\documentstyle[amssymbols,A4]{article}
\def\model{${\cal B}_{3D}$}
\def\a{\alpha}
\def\|#1>{\vert#1\rangle}

\begin{document}
\begin{titlepage}
\noindent
\hfill 
	{PAR--LPTHE 92--16}\par
\hfill\today\par
\vskip 2cm
\centerline{\bf TOWARDS THREE-DIMENSIONAL BETHE ANSATZ}
\vskip 3cm
\begin{center}
\obeylines
M. Bellon,  S. Boukraa, J-M. Maillard, C-M. Viallet
Laboratoire de Physique Th\'eorique et des Hautes Energies
Unit\'e associ\'ee au C.N.R.S ( URA 280)
Universit\'e de Paris 6--Paris 7, Tour 16, $1^{\rm er}$ \'etage, bo\^\i te 126
4 Place Jussieu/ F--75252 PARIS Cedex 05
\end{center}

\vskip 3cm
\begin{abstract}

We introduce a ``pre-Bethe-Ansatz'' system of equations for three
dimensional vertex models.  We bring to the light various algebraic curves of
high genus and discuss some situations where these curves simplify.  As a
result we describe remarkable subvarieties of the space of
parameters.

\end{abstract}

\noindent
 {\bf AMS Classification scheme numbers}: 82A68, 82A69, 14E05, 14J50,
16A46, 16A24, 11D41

\vfill
\centerline{work supported by CNRS}

\end{titlepage}

\def\C{{\Bbb C}}
\def\Ip{{\cal I}_{\rm proj}}
\def\CP{{\Bbb CP}}

\section{Introduction}

The purpose of this paper is to sketch how ideas introduced in the study of
the sixteen vertex model in~\cite{prl2} can  be generalized to
higher lattice dimensions.  We think that the ideas developed here are
relevant tools for the analysis of lattice models in three or more
dimensions, a widely unexplored area.

In this paper we first recall the basic results obtained for the
two-dimensional case, with a special emphasis on the symmetries they uncover.
We then introduce the simplest three-dimensional generalization of these
results.  In order to be more concrete, we describe a specific model which
naturally generalizes the Baxter model.  Finally, we show how this general
construction points to a number of algebraic varieties of interest.


\section{Preliminaries to  the  Bethe Ansatz}
\label{towards}

One of the keys to the Bethe Ansatz for two-dimensional vertex models
is the existence (see for instance equations (B.10), (B.11a)
in~\cite{Ba73}) of vectors which are pure tensor products (of the form
$v \otimes w$), and  which the $R$-matrix maps onto another pure tensor
product $v' \otimes w'$~\cite{Krixx}.  The ``pre-Bethe Ansatz''
equation is therefore:
\begin{equation}
\label{eqt}
	 R \; ( v \otimes w ) = \mu \; v'\otimes w' ,
\end{equation}
A group $G_{Bethe} \simeq sl_2 \times  sl_2 \times sl_2 \times sl_2 $
naturally acts on this equation, extending the weak-graph
transformations~\cite{prl2}.  With the following notations
\begin{equation}
	v=\pmatrix{ 1 \cr p \cr}, \quad w=\pmatrix{ 1 \cr q \cr}, \quad
	v'=\pmatrix{ 1 \cr p' \cr}, \quad w'=\pmatrix{ 1 \cr q' \cr},
\end{equation}
the elimination of $w$ and $w'$ from (\ref{eqt}) yields  the biquadratic
relation:
\begin{equation}
\label{biq1}
	\alpha_{00} + \alpha_{10} \; p + \alpha_{01} \; p'
		+ \alpha_{20} \; p^2 + \alpha_{02} \; p'^2 + 2 \alpha_{11} \; p p'
		+ \alpha_{21} \; p^2 p' + \alpha_{12} \; p p'^2
		+ \alpha_{22} \; p^2 p'^2 =0 .
\end{equation}
The $\alpha_{ij}$'s are quadratic polynomials in the homogeneous
parameters $a_1, \cdots, d_4$ of the $R$-matrix.

Equation (\ref{biq1}) (and the similar one for $q$ and $q'$) is
a {\it quadratic Frobenius relation}~\cite{Ga83,frobenius,Mu84} on theta
functions. Let us recall that the Frobenius relations are intertwining
relations of the product of two theta functions~\cite{Ga83}.  The
``intertwining'' matrix $R$ is expressible in terms of ratio and product of
theta functions depending on the {\it difference} of the arguments (spectral
parameters) of the theta functions appearing in the parametrization of $p$,
$p'$ (resp.\ $q$, $q'$)~\cite{Mu84,Ga83}.  This intertwining functional
relations on theta functions enable to represent a Zamolodchikov algebra
(see~\cite{ZaZa79}) which is ``almost''%
	\footnote{The very existence of a Zamolodchikov algebra implies
	the Yang-Baxter equations to be satisfied {\it provided} the
	linear independence of the expressions ${\cal A}_i(\theta_1){\cal A}_j
	(\theta_2){\cal A}_k(\theta_3)$.  This independence condition
	is apparently not satisfied for theta functions of several
	parameters~\cite{Sch82}.}
a sufficient condition for the Yang-Baxter equations to be satisfied:
\begin{equation}
	{\cal A}_i(\theta_1)\;{\cal A}_j(\theta_2) = \sum_{kl} R^{kl}_{ij}
		(\theta_1-\theta_2)\; {\cal A}_k(\theta_2)\;{\cal A}_l(\theta_1)
\end{equation}

In~\cite{prl2}, it has been shown that equation (\ref{biq1}) defines an
elliptic curve in the $p,p'$ plane.  The discriminant of (\ref{biq1}) seen
as a quadratic polynomial in $p$ is:
\begin{equation}
\label{del}
	D= \alpha p'^4 + 4 \beta p'^3 + 6 \gamma p'^2 + 4\beta' p' +\alpha' .
\end{equation}
The transformations of $D$ under the $sl_2$ group acting on $v'$
(homographic transformations of $p'$) yield two fundamental invariants
$g_2$ and $g_3$~\cite{Di90,He1854} and a {\it modular invariant}
$J=g_3^2/(g_2^3 - 27 g_3^2)$:
\begin{eqnarray}
	g_2 &=& \alpha \alpha' - 4 \beta \beta' + 3 \gamma^2 \\
	g_3 &=& \alpha \gamma \alpha' + 2 \beta \gamma \beta'
		- \alpha \beta'^2-\alpha' \beta^2 - \gamma^3
\end{eqnarray}

Exchanging  the role of $p$ and $p'$ in equation (\ref{biq1}) leads to the
same $g_2$, $g_3$ and $J$.  More remarkably, the same operations applied to
the equation in $q$ and $q'$ similar to (\ref{biq1}), lead to {\it the same
$g_2$ and $g_3$}. This is a quite non-trivial result.  One can therefore
associate {\it the same Weierstrass's canonical form} (elliptic curve),
$y^2=4x^3-g_2x-g_3$, to equation (\ref{biq1}), the similar equation in $q$
and $q'$ and also to the   elliptic curve, intersection of quadrics in
${\C}{\bf P}_{15}$, which corresponds to the orbits of $\Gamma$ (see
Figure~1 in~\cite{prl2}).

The algebraic variety defined by  $J= \infty$, i.e. $g_2^3 -27\;g_3^2 =0$,
is a remarkable variety in the parameter space ${\C}{\bf P}_{15}$: {\it the
elliptic parametrization of (\ref{biq1}) becomes  a rational one}.
Recalling the well-known example of the symmetric eight-vertex Baxter
model~\cite{Ba72,Ba73,Ba78}, this reduction to a rational parametrization
corresponds to two remarkable situations in the parameter space: the {\it
disorder varieties}, for which some dimensional reduction
occurs~\cite{JaMa85}, and the {\it critical varieties}~\cite{Ba81,prl2}.
Quite surprisingly, these two varieties {\it of a priori different nature
appear here on the same footing}.  It would be pointless to write down the
condition $J=\infty$, since it is a sum of several millions of monomials of
degree 24 in the sixteen homogeneous parameters of the sixteen-vertex
model.

Noticeably, the Frobenius relations do exist for theta functions of several
variables ($g$ variables) leading to some associated Zamolodchikov
algebra~\cite{Ga83}.  However, they do not lead to Yang-Baxter
equations~\cite{Sch82}.  Clearly, there is room for models satisfying a
non-trivial functional relation like equation (\ref{eqt}) (which is a key
for the explicit construction of the Bethe Ansatz) but not the Yang-Baxter
equations.  We may for instance consider the generalization of the symmetric
eight-vertex model constructed in~\cite{Sch82} for which there exists a
parametrization in terms of theta functions of $g$ variables: this model can
be seen as a coupling of $g$ replicas of the Baxter model, the arrows of the
vertex  taking values in ${\bf Z}_2^g$ and the row-to-row transfer matrix
(with periodic boundary conditions) being a $2^{gN} \times 2^{gN}$ matrix.
If the Yang-Baxter equations were satisfied (which is apparently not the
case~\cite{Sch82}), they would yield the existence of a ($g$-dimensional)
family of commuting transfer matrices.  This is certainly too demanding.
The existence of transfer matrices commuting {\it only} in a {\it subspace}
of the $2^{gN}$ vector space, corresponds to interesting models and may be
sufficient to calculate their partition function (largest eigenvalue of the
transfer matrix).  One must recall the example of the disorder
solutions~\cite{JaMa85,HaJoMa87}, for which a commutation of the (diagonal)
transfer matrices is satisfied on a one-dimensional vector space, which is
actually sufficient to calculate the partition function (see~\cite{HaJoMa87}
and page 247 of~\cite{Ba81}).

The exchange of $p$ and~$p'$ (resp.\ $q$ and~$q'$), as well as the symmetries
described in~\cite{bmv2b}, are realized as {\it shifts
of the spectral parameter $\theta$}  of the elliptic curve (see for
instance~\cite{Ka74}).  This is a  key point for the explicit construction
of the Bethe Ansatz, since it enables to build a vector
\begin{equation}
\label{prod}
	\| \Psi (\theta) >  = \bigotimes _n e(\theta+2 \eta n),
\end{equation}
which transforms very simply by the row-to-row transfer matrix:
\begin{equation}
  \| \Psi (\theta)  >  \rightarrow  \| \Psi (\theta+2 \eta ) >.
\end{equation}

When acting on $\Psi$, the translation by  a lattice spacing amounts to the
{\em shift} of the spectral parameter $\theta$.  Taking into account the
translation invariance of the lattice with periodic boundary conditions, one
can ``Fourier transform'' $\| \Psi (\theta) >$ and get an eigenvector
$\| \widehat{\Psi} (\theta) >$ of the transfer matrix (see p.~148
of~\cite{Ka74}).  Notice that this construction works for theta functions of
$g$ variables using the corresponding Frobenius relations~\cite{Mu84}. One
therefore has a commutation of transfer matrices on this (one dimensional !)
vector space.  This is just the first step for the construction of the Bethe
Ansatz. The second step amounts to build eigenvectors made up of product
vectors (similar to (\ref{prod})) with a given number of ``deviating''
factors (see p.  148--150 of~\cite{Ka74} and ~\cite{Ba73}).

\section{Towards three-dimensional Bethe Ansatz}
A similar analysis can be performed in higher dimensions.  In this
section, we show how we can generalize these results in dimension three.
In a specific example, we find three curves linked to a matrix $R$ living on
a higher dimensional variety~\cite{prl}.

We denote  $w(i,j,k,l,m,n)$ the Boltzmann weight of a given three-dimensional
vertex.  We shall only consider the simplest case where each of the spins $i$,
$j$, $k$, $l$, $m$ and~$n$ can take only two values.  The vertex  weights may
be arranged in an $8 \times 8$ matrix of entries:
\begin{equation}
        R^{ijk}_{lmn} = w (i,j,k,l,m,n).
\label{3R}
\end{equation}

The natural generalization of the ``pre-Bethe Ansatz'' equation (\ref{eqt}) is:
\begin{equation}
\label{pt}
 R \; ( u \otimes v \otimes w ) = \mu \cdot u' \otimes v'\otimes w' .
\end{equation}
Let us introduce the notation:
\begin{equation}
	u=\pmatrix{ 1 \cr p \cr}, \;\; v=\pmatrix{ 1 \cr q \cr},
	\;\; w=\pmatrix{ 1 \cr r \cr}, \;\; u'=\pmatrix{ 1 \cr p' \cr},
	\;\; v'=\pmatrix{ 1 \cr q' \cr}, \;\; w'=\pmatrix{ 1 \cr r' \cr}.
\end{equation}
In the following subsections, we shall recall the symmetries of a three
dimensional vertex model, as described in~\cite{bmv2b}.

\subsection{The group of inversions $\Gamma_{3D}$}

As in~\cite{bmv2b}, we first introduce the involution  $I$ changing $R$ to
its matrix inverse (we let appear an overall factor $\lambda$ since the entry
of $R$ are taken projectively).
\begin{equation}
\label{Itetra}
        \sum_{\a_1,\a_2,\a_3} (IR)_{\a_1\a_2\a_3}^{i_1i_2i_3} \cdot
                R^{\a_1\a_2\a_3}_{j_1j_2j_3} = \lambda \; \delta^{i_1}_{j_1} \;
                        \delta^{i_2}_{j_2} \; \delta^{i_3}_{j_3} .
\end{equation}
Multiplying both sides of (\ref{pt}) by $IR$, we get an equation of the same
form as (\ref{pt}) with $u$ and $u'$, $v$ and $v'$ and $w$ and $w'$ exchanged
and $R$ replaced by $IR$.

In~\cite{bmv2b}, we also introduced three partial transposition $t_1$, $t_2$
and~$t_3$.  $t_1$ is defined by:
\begin{equation}
\label{trp}
        (t_1R)^{i_1i_2i_3}_{j_1j_2j_3} = R^{j_1i_2i_3}_{i_1j_2j_3}.
\end{equation}
The definitions of $t_2$ and $t_3$ are similar.

The four involutions $I$ and $t_i$ ($i=1$, 2, 3) generate an infinite discrete
group $\Gamma_{3D}$~\cite{bmv2b}.  The so-called inversion relations of
the statistical mechanics model can be simply expressed with these building
blocks.  They are:
\begin{equation}
\label{trq}
        I,\quad J=t_1It_2t_3, \quad K=t_2It_3t_1, \quad L=t_3It_1t_2.
\end{equation}
Considering the parameter space as a projective space (the entries of
the $R$-matrix are homogeneous parameters), the elements of the group
$\Gamma_{3D}$ have a {\it non-linear } representation in terms of {\it
birational transformations}.  This group of symmetry of the parameter
space of the model is very large.  The number of elements of length $l$
grows exponentially with $l$.  It is actually  a {\em hyperbolic}
Coxeter group~\cite{Hu90}.  The symmetry group of the Yang-Baxter equations
in two dimensions is a mere affine Coxeter group~\cite{Hu90,bmv2,bmv2b}.

The group  $\Gamma_{3D}$ has been shown in~\cite{bmv2b} to enter the
description of the group of automorphisms of the tetrahedron equations
(generalization of the Yang-Baxter equations in three dimensions).
We shall use this symmetry group beyond integrability, that is to say for
models which do not have to verify the tetrahedron equations.

\subsection{Weak-graph duality for 3D models: the gauge group ${G}$}

A ``gauge'' group ${G}= sl_2 \times sl_2 \times sl_2$ acts {\it
linearly} on the matrix $R$ by similarity transformations
(the weak-graph transformations, see~\cite{GaHi75} for details).  If
$g=g_1 \times g_2 \times g_3$, we define:
\begin{equation}
	g(R)= g_1\;  g_2\;  g_3 \cdot R\cdot g_1^{-1}g_2^{-1}g_3^{-1} .
\label{gauge}
\end{equation}
Each of the $g_i$'s acts on the corresponding vector space and $g_1$ for
example is a short hand notation for $g_1\otimes {\Bbb I}\otimes {\Bbb I}$.
The action of ${G}$ and $\Gamma_{3D}$ do not commute.  However, $G$ and $I$ do
commute, and the commutation relation between the $t_i$'s and $G$ gives a
rather simple semi-direct product structure to the combined group:
\begin{eqnarray}
	t_1 g &=& g^{t_1} t_1,\\
\noalign{\noindent with:}
	g^{t_1} &=& {}^t\!g_1^{-1} \times g_2 \times g_3,
\end{eqnarray}
and similar relations for $t_2$ and $t_3$.  In particular, $\Gamma_{3D}$
sends orbits of $G$ onto orbits of  $G$.  The compatibility of these two
groups is described in~\cite{prl2} in a two-dimensional case, the
sixteen-vertex model.

The effect of such a transformation on the pre-Bethe-Ansatz equation
(\ref{pt}) is simple: $g_1$ act naturally on $u$ and $u'$, $g_2$ on $v$ and
$v'$ and $g_3$ on $w$ and $w'$.

\section{A three-dimensional model} \label{tr}

The most general vertex models on a cubic lattice has a large number
of  parameters~(sixty-four).  We therefore impose some relations on the
Boltzmann weights  of the three-dimensional vertex.  We require that these
relations are invariant under the inverse $I$~\cite{bmv1,bmv4} and the three
partial transpositions $t_1$, $t_2$ and $t_3$ (equation (\ref{trp})).  They
will thus be invariant under the group  $\Gamma_{3D}$.  We are particularly
interested in generalizations of the Baxter symmetric eight-vertex model,
and define here a specific three-dimensional model, denoted in the sequel
\model.  It is possible to ``project'' down a three-dimensional model onto a
bidimensional one by just taking the trace of the matrix $R$ on one of the
spaces 1,2, or 3: take for example space 3.
\begin{equation}
	{\widehat R}^{ij}_{kl} = \sum_{\alpha_3} R^{i,j,\alpha_3}_{k,l,\alpha_3}
\label{proj3}
\end{equation}
The constraints verified by \model\ are such that the three possible
projections are symmetric Baxter models.

We define \model\ by  imposing the following restrictions
on the entries~\cite{bmv2b}:
\begin{eqnarray}
\label{v1}
        R^{i_1 i_2 i_3}_{j_1 j_2 j_3}&=&R^{-i_1,-i_2,- i_3}_{-j_1,- j_2,-
j_3}\\
\label{v2}
        R^{i_1 i_2 i_3}_{j_1 j_2 j_3}&=&0 \quad \mbox{ if }\quad
                {i_1 i_2 i_3}{j_1 j_2 j_3}=-1
\end{eqnarray}
These constraints imply that the $8\times8$ matrix $R$ is
the direct product of two times the same $4\times4$ submatrix~\cite{prl}.
It is further possible to impose that this $4 \times 4$ matrix is symmetric,
since such a symmetry is preserved by the partial transpositions
$t_1$, $ t_2$, $ t_3$~\cite{bmv2b,prl}, that is:
\begin{equation}
\label{v3}
	 R^{i_1 i_2 i_3}_{j_1 j_2 j_3} =  R^{j_1 j_2 j_3}_{i_1 i_2 i_3}
\end{equation}

We shall use  the following notations for the entries of this
$4\times4$ submatrix:
\begin{equation}
        \left( \begin{array}{cccc}
                                a   & d_1 & d_2 & d_3 \\
                                d_1 & b_1 & c_3 & c_2 \\
                                d_2 & c_3 & b_2 & c_1 \\
                                d_3 & c_2 & c_1 & b_3
        \end{array} \right).
\label{M1}
\end{equation}
The four rows and columns of this matrix correspond to the states
$(+,+,+)$, $(+,-,-)$, $(-,+,-)$ and~$(-,-,+)$ of the triplets
$(i_1,i_2,i_3)$ or $(j_1,j_2, j_3)$. The
matrix $R$ can be completed by spin reversal, according to
(\ref{v1}).  Note that $t_1$ (resp.\ $t_2$, $t_3$) simply exchanges
$c_2$ with  $d_2$ and  $c_3$ with $d_3$ (resp. circular permutations).
$I$ acts as the inversion of this $4\times4$ matrix.


It is quite remarkable that there exist four quantities which are
covariant by all the four generating involutions $I$, $t_1$, $t_2$, $t_3$,
and therefore the whole group $\Gamma_{3D}$.  Let us introduce:
\begin{equation}
\label{pols}
        p_3 = a b_3 + b_1 b_2 - c_3^2 -d_3^2, \qquad
        q_3 = c_1 d_1 - c_2 d_2,
\end{equation}
and the polynomials obtained by permutations of 1, 2 and~3. They form a
five dimensional space of polynomials. Any ratio of these
polynomials is invariant under all the four generating
involutions $I$, $t_1$, $t_2$, $t_3$. ${\C}{\bf P}_9$ is thus foliated by
five dimensional algebraic varieties invariant under the {\it
whole group} $\Gamma_{3D}$.  We can also express it by saying that the
polynomials (\ref{pols}) define a map from the parameter space
$\C{\bf P}_9$ to $\C{\bf P}_4$ invariant under $\Gamma_{3D}$.

If we consider a subgroup $\Gamma_2$ generated by {\it only}
two involutions, say $I$ and $L$ (\ref{trq}) or equivalently $I$ and $t_3$, one
gets {\it three  more} independent covariant polynomials leading to algebraic
surfaces (see Fig. 1).  They read:
\begin{eqnarray}
	r_3 &=& a b_3 - b_1 b_2 - c_3^2 +d_3^2,\nonumber\\
	s_3 &=& (a+b_3) c_3 -d_1 d_2 -c_1 c_2, \nonumber\\
	t_3 &=& (b_1 + b_2) d_3 - d_1 c_2 - c_1 d_2.
\label{more}
\end{eqnarray}

 From the projection (\ref{proj3}) we get a Baxter model.  If we denote by
$a_B$, $b_B$, $c_B$, $d_B$ the non zero entries of the $R$-matrix of this
model, we have:
\begin{equation}
\label{corres}
	a_B= a+b_3, \qquad b_B = b_1+b_2,
	\qquad c_B = 2c_3, \qquad d_B=  2d_3 .
\end{equation}

\section{Study of the three-dimensional ``pre-Bethe'' equations.}

\subsection{A first attempt.}

In the study of the equation (\ref{pt}), we can start by eliminating the
variables $p$ and $p'$.  We obtain the following system of five equations on
the remaining variables $q$, $q'$, $r$ and~$r'$:
\begin{eqnarray}
	0 &=& c_3 d_3
	- b_2 b_3 \;q^2
	- c_1^2 \;r^2
	+ ( d_2 d_3- b_2 c_1+ c_2 c_3- b_3 c_1) \;rq
	+ c_2 d_2 \;q^2r^2
	-a b_1  \;q'^2
\nonumber \\ && {}
	+ (a b_3- c_3^2+ b_1 b_2- d_3^ 2 )  \;qq'
	+ (a c_1 - c_2 c_3+ b_1 c_1- d_2 d_3 )  \;rq'
\nonumber \\ && {}
	+ ( b_2 d_1- c_3 d_2- c_2 d_3+ b_3 d_1 )  \;rq^2q'
	+ 2( c_1 d_1- c_2 d_2 )  \;r^2qq'
\nonumber \\ && {}
	- (a d_1+ b_1 d_1 - c_2 d_3- c_3 d_2)  \;rqq'^2
	+ c_3 d_3  \;q^2 q'^2
	+ c_2 d_2 \;r ^2 q'^2
	- d_1^2 \;q^2r^2 q'^2,
\\	\noalign{\medskip}                      
	0 &=& ( c_1 c_3- b_2 c_2) \;q
	+ ( b_3 c_3- c_1 c_2 ) \;r
	+ ( c_1 d_2- b_2 d_3 ) \;q^2r
	+ ( b_3 d_2- c_1 d_3 ) \;qr^2
\nonumber \\ && {}
	+a c_2  \;q'
	-a c_3  \;r'
	- c_1 d_3 \;q^2 q'
	- b_3 d_2 \;r^2 q'
\nonumber \\ && {}
	+ b_2 d_3 \;q^2 r'
	+ c_1 d_2 \;r^2 r'
	+ d_1 d_3 \;q^2r^2 q'
	- d_1 d_2 \;q^2r^2 r'
\nonumber \\ && {}
	+ (a d_3+ c_2 d_1 - b_3 d_3- c_1 d_2)  \;qrq'
	- (a d_2+ c_3 d_1- b_2 d_2- c_1 d_3 )  \;qr r' ,
\\	\noalign{\medskip}			
	0 &=& c_3 d_2
	- b_2 c_1 \;(q^2 +r^2)
	+ ( c_3^2- b_2^2+ d_2^2- c_1^2 ) \;rq
	+ (a b_2- d_2^2 )  \;rq'
	+ (a c_1- d_2 d_3 )  \;qq'
\nonumber \\ && {}
	+ ( b_1 b_2- c_3^2 ) \;q r'
	-a b_1  \;q' r'
	+ c_3 d_2 \;q^2r^2
	+ ( b_2 d_1- c_3 d_2 )  \;r^2qq'
	+ ( c_1 d_1- c_3 d_3 )  \;r q^2q'
\nonumber \\ && {}
	+ ( c_1 d_1- c_2 d_2)  \;r^2qr'
	- d_1^2 \;q^2r^2 q' r'
	+ ( c_2 d_3+ c_3 d_2-a d_1- b_1 d_1 )  \;rqr' q'
\nonumber \\ && {}
	+ ( b_1 c_1- c_2 c_3)  \;rr'
	+ c_2 d_2 \;r^2 q' r'
	+ ( b_2 d_1- c_3 d_2 )  \;r'rq^2
	+ c_3 d_3 \;q^2 q' r',
\\	\noalign{\medbreak}			
	0 &=& c_3 d_1
	- b_2 c_2 \;q^2
	- c_1 c_3 \;r^2
	+ ( d_1 d_2- c_1 c_2- b_2 c_3+ b_1 c_3 ) \;rq
	+ b_1 d_2 \;q^2r^2
\nonumber \\ && {}
	+ (a c_2- d_1 d_3 )  \;qq'
	+ (a c_3- d_1 d_2 )  \;rq'
	-a c_3  \;q' r'
\nonumber \\ && {}
	+ ( c_3 d_1- b_1 d_2 )  \;r^2qq'
	+ ( c_2 d_1- b_1 d_3 )  \;rq^2q'
	+ c_1 d_2 \;r^2 q' r'
	+ b_2 d_3 \;q^2 q' r'
\nonumber \\ && {}
	- (a d_2+ c_3 d_1- c_1 d_3- b_2 d_2 )  \;rqr' q'
	- d_1 d_2 \;q^2r^2 q' r',
 \\	\noalign{\medskip}			
	0 &=& ( c_3 d_2- b_2 d_1 ) \;q
	+ ( c_3 d_3 - c_1 d_1 ) \;r
	- (a b_2 - d_2^2) \;q^2r
	- (a c_1 - d_2 d_3) \;qr^2
	+a d_1  \;q'
\nonumber \\ && {}
	+ (a^2 + d_1^2- d_3^2- d_2^2)  \;rqq'
	- d_2d_3 \;(q^2 + r^2) q'
	+a d_1 \;q^2r^2 q'
	+ ( b_1 b_2 - c_3^2 ) \;q q' r'
\nonumber \\ && {}
	+ ( b_1 c_1- c_2 c_3 )  \;rr' q'
	- (a d_1+ b_1 d_1 -c_2 d_3- c_3 d_2)  \;rqr' q'^2
	+ c_2 d_2 \;r^2 q'^2 r'
	-a b_1  \;q'^2 r'
\nonumber \\ && {}
	+ ( c_1 d_1- c_2 d_2 )  \;r^2qr' q'
	+ ( b_2 d_1- c_3 d_2)  \;rq^2r' q'
	- d_1^2 \;q^2r^2 q'^2 r'
	+ c_3 d_3 \;q^2 q'^2 r'
{}.
\end{eqnarray}
Two similar system of equations can be obtained by the elimination of the
pair of variables $q$ and~$q'$ or $r$ and~$r'$.  The equations are of degree
two in each of the variables, that is a overall maximum degree of eight.  In
fact only one is of degree seven, three are of degree six and one is of
degree five.

The only apparent property of this system is the
invariance by changing the sign of each of the variables $q$, $q'$, $r$
and~$r'$, which is linked to the zeroes of the $R$ matrix for $ijklmn=-1$.

The spin reversal symmetry of the $R$ matrix and the change of $R$ into its
inverse $IR$ have no visible consequences.  This is due to
the particular choice made in the elimination of $p$ and~$p'$.  The equations
we have written are just five out of
a system of thirty-six equations with a number of relations among them.  On
this full system, the symmetries should be more manifest.

\subsection{Necessary conditions for \model.}
Since this direct attempt to find a full solution to (\ref{pt}) leads to such
a confuse result, we shall a study some necessary conditions.  We replace the
tensor product $u\otimes v$ and $u' \otimes v'$ by general vectors in the
tensor product of space 1 and~2 $U$ and~$U'$.  (\ref{pt}) becomes:
\begin{equation}
	R\; (U\otimes w) = \mu U'\otimes w' .
\label{ptm}
\end{equation}
This can be written:
\begin{equation}
\pmatrix{                                R_1  & R_2  \cr
                                         R_3  & R_4  \cr           }
\pmatrix{ U \cr rU \cr}
=\mu \pmatrix{ U' \cr r'U' \cr},
\end{equation}
with  $R_1, \cdots, R_4$ the $4\times4$ blocs of $R$.

Eliminating  $U$ and $U'$  gives the {\it  necessary} condition:
\begin{equation}
\label{v4}
det(R_1 \; r' +R_2 \; r r' - R_3 -R_4 \; r)=0 .
\end{equation}
For the model introduced in section (\ref{tr}) (eqs. (\ref{v1}), (\ref{v2}),
(\ref{v3})), equation (\ref{v4}) reads in terms of $r$ and $r'$:
\begin{eqnarray}
\label{can}
A_3 \cdot ( r^4 r'^4 +1) + B_3  \cdot ( r^4 r'^2 + r^2 r'^4 + r^2 +r'^2 )
	+C_3  \cdot (r^4 + r'^4) \quad \nonumber \\
+D_3  \cdot (r^3 r'^3 +r r')+E_3  \cdot (r^3 r' + r r'^3) +F_3  \cdot r^2 r'^2
	&=& 0  .
\end{eqnarray}
Here $A_3, \dots, F_3$ are polynomial expressions of degree four in the
homogeneous entries of the $R$-matrix (\ref{M1}) $(a, \cdots, d_3)$.
The simplest expressions are:
\begin{eqnarray*}
	A_3&=& (c_1 d_1-c_2 d_2)^2 = q_3^2,\\
	C_3&=&  (b_1 b_2 - c_3^2)( a b_3 - d_3^2) =
		{\textstyle1\over4}(p_3+r_3)(p_3-r_3).
\end{eqnarray*}
In fact, all the coefficients $A_3, \cdots, F_3$ can be expressed as
quadratic expressions in the polynomials invariant by the subgroup $\Gamma_2$
of $\Gamma_{3D}$ listed in (\ref{pols}) and (\ref{more}).  This shows that
equation (\ref{can}) is invariant by this infinite group.
Moreover, they verify the relation:
\begin{equation}
\label{rela}
4 A_3 \cdot (F_3-2 A_3 + 2 C_3 + 2 E_3) = (D_3 + 2 B_3 )^2  .
\end{equation}
Relation (\ref{rela}) is actually the condition for relation (\ref{can}) to
become, when $r=r'$, the square of
\begin{equation}
	  (c_2 d_2-c_1 d_1)(r^4+1) + r^2
		(ab_2+b_1 b_3-c_2^2-d_2^2-ab_1-b_2 b_3+c_1^2+d_1^2).
\end{equation}
We recognize $q_3$ in the coefficient of $r^4+1$ and $p_2-p_1$ (\ref{pols})
in the coefficient of $r^2$.  These coefficients are thus
$\Gamma_{3D}$-covariant  polynomials.


Of course two similar eliminations can be performed on (\ref{pt}) yielding
constraints between $p$ and $p'$  (resp. $q$ and $q'$).

To take into account the  symmetry of (\ref{can}) by the exchanges
$r \leftrightarrow r'$, $r \leftrightarrow 1/r$
and $r' \leftrightarrow 1/r'$, one may  introduce the variables $X= rr'+ 1/rr'$
and $Y=r/r'+r'/r$. Equation (\ref{can}) then becomes a {\it conic}:
\begin{equation}
\label{conic}
	A_3\; X^2 + B_3 \;XY + C_3\; Y^2 +D_3\; X +E_3\;Y +\widetilde{F}_3 =0,
\end{equation}
with $\widetilde{F}_3= F_3 -2 A_3 -2 C_3$.

An invariant $\Ip$ is naturally associated to the conic (\ref{conic}), it is
the determinant of the $3 \times 3$ matrix $M$~\cite{Di90}:
\begin{equation}
\label{M}
	M = \pmatrix{
		A_3	&  B_3/2  	&  D_3/2	\cr
		B_3/2	&  C_3	&  E_3/2	\cr
		D_3/2	&  E_3/2  	&  \widetilde{F}_3  \cr}
\end{equation}
The value of this  invariant is (taking into account the relations between
the entries of $M$):
$\Ip=-{\cal I}^2/4A_3$, with
\begin{equation}
\label{I}
	{\cal I}=2\,B_3^{2}+D_3B_3-2\,E_3A_3-8\,A_3C_3.
\end{equation}
$\Ip=0$ is a projectively invariant condition for the conic,  meaning  that it
 is the union of two lines. Remark that this does not imply the existence
of a rational uniformization of (\ref{can}).  A similar phenomenon happens in
the Baxter model, for which (\ref{biq1}) becomes   linear in $X$ and $Y$, but
this does not provide  the  elliptic parametrization.

In order to obtain a parametrization of  (\ref{can}), we
look at it  as a  polynomial of degree four in $r$, the
coefficients being polynomials in $r'$:
\begin{equation}
\label{u1}
	\alpha (r')  r^4 + 4  \beta (r')  r^3 + 6  \gamma (r')  r^2
		+ 4 \beta '(r')  r+ \alpha '(r') =0
\end{equation}
Its discriminant  reads~\cite{Di90,He1854}:
\begin{eqnarray}
\label{u2}
	\Delta (r') &=& g_2(r')^3-27\,g_3(r')^2,\\
\noalign{\noindent{with:}}
\label{u3}
	g_2(r')&=& \alpha(r') \alpha '(r')-4\, \beta(r') \beta '(r')
		+3\,\gamma(r')^2,\\
\noalign{\noindent{and:}}
\label{u4}
	g_3(r')&=& \alpha(r') \gamma(r') \alpha '(r')
		+2\, \beta(r') \gamma(r') \beta '(r') \nonumber\\
		&& - \alpha(r') \beta '(r')^2
		- \beta(r')^2 \alpha '(r')- \gamma(r')^3.
\end{eqnarray}

For a general vertex model in three dimensions, equation (\ref{v4})
leads to an equation like (\ref{u1}) where $\alpha(r'),\; \beta(r'),\;
\gamma(r'),\; \beta'(r'),\; \alpha'(r')$ are polynomials of degree four
in $r'$. The polynomials  $g_2(r'), \; g_3(r')$, and  $\Delta(r')$ are
polynomials of
degree 8, 12 and 24 respectively. For the  general vertex model the
hyperelliptic curve $y^2=\Delta(r')$ is a genus eleven curve.

However, one verifies easily that, for the model \model,
  $\Delta(r')$  is a polynomial of degree twelve in $r'^2$.
Moreover  the polynomial $\Delta(r')/r'^{12}$ is symmetric under
the inversion $r' \leftrightarrow 1/r'$. Hence, introducing the
variable $s'=r'^{2}+r'^{-2}$, $\Delta(r')/r'^{12}$ becomes a {\it
degree six} polynomial in $s'$, denoted $P_6(s')$.

If $P_6(s')$ were a generic polynomial of degree six, the
hyperelliptic curve $y^2=P_6(s')$ would be a genus two curve, meaning
that, as far as parametrization is concerned, one is obliged to deal
with theta functions of two variables (the Jacobian associated to the
genus two curve~\cite{Mu84}) or automorphic functions (see~\cite{Sh77} and
page 455 of ~\cite{WhitWat}).  One can envisage handy parametrization when
the hyperelliptic curves degenerate into elliptic ones, that is, when
two roots of $P_6(s')$ coincide, or equivalently when the discriminant
of $P_6(s')$ vanishes\footnote{
	One should note that this is just an auxiliary parametrization and not a
	uniformization of eq.~(\ref{u1}).
}.

It is important to note that the model of section (\ref{tr}), {\it
corresponds to such a situation} where $P_6(s')$ can be written as:
\begin{equation}
P_6(s')=  (s'-s_0 )^{2} \cdot P_4(s')
\label{p6p4}
\end{equation}
where $P_4(s')$ is a polynomial of the fourth degree in $s'$,
containing 289 monomials of degree eight in the coefficients $A_3,
\cdots, E_3$.  Noticeably, $s_0$ is a quite simple expression:
\begin{equation}
 s_0=-{\frac {D_3+2\,B_3}{2\,A_3}}
\end{equation}
which reads in terms of the entries $a, \cdots, d_3$ of $R$:
\begin{equation}
\label{tt}
 s_0={\frac { \left ( c_1^{2}-c_2^{2}+d_1^{2}-d_2^{2}\right )  -\left
(a-b_3\right )
\left (b_1-b_2\right )}{c_1\,d_1-c_2\,d_2}}  \label{sols}
\end{equation}
Expression (\ref{tt}) is  invariant under the group
$\Gamma_{3D}$ (see equations (\ref{pols})). In equation (\ref{can}),
index 3, and the equations similar to (\ref{can})
relating $p$ and $p'$, or $q$ and $q'$, lead to   equations like
(\ref{sols}),  where 1,2 and 3 are permuted.
  It would be
interesting to look  for conditions on the entries of the
$R$  such that these  three elliptic curves identify.


Clearly, a particular variety plays a special role: the subvariety
 in the space of models where the three elliptic curves reduce
  to rational ones.
This algebraic variety is a good candidate for being a set of  critical points
(or disorder points) for \model\ though it is only  a codimension-three
subvariety of the
codimension-one critical manifold we are looking for.

To sum up, the three-dimensional vertex model \model\ yields a
generalization of the intertwining
quadratic Frobenius relations in the form of an intertwining of three
different elliptic curves by an $R$-matrix living on an algebraic
variety of dimension five given by the intersection of five quadrics
(equations (\ref{pols})).

\subsection{Further analysis}
The polynomial $P_4(s')$  appearing in equation (\ref{p6p4}) is worth
analyzing.
We see that  $g_2$ and $\;\Delta=g_2^3-27 g_3^2$ factorize:
\begin{equation}
	g_2 = A_3^3 \cdot g_2^{(1)} \cdot  g_2^{(2)}, \qquad
	g_3 = A_3^4 \cdot g_3^{(1)}, \qquad
	\Delta = A_3^8 \cdot \Delta_1 \Delta_2 \Delta_3 \Delta_4 \Delta_5^3
\end{equation}
with:
\begin{eqnarray*}
	g_2^{(1)}&=& 3\,E_3^{2}A_3 +8\,C_3E_3A_3 +16\,A_3^{2}C_3 -
		D_3^{2}C_3 - 16\,C_3^{2}A_3 - E_3B_3D_3   \\
	&& -2\,B_3^{2}E_3 - 4\,B_3^{2}A_3 + 4\,C_3B_3^{2},   \\
	\Delta_1&=& 2\,E_3A_3 - D_3B_3, \qquad
	\Delta_2= 2\,B_3^{2}+D_3B_3-2\,E_3A_3-8\,A_3C_3, \\
	\Delta_3&=&
		4\,B_3^{2}A_3 + 16\,C_3^{2}A_3-4\,C_3B_3^{2}-D_3^{2}C_3
		+ E_3^{2}A_3 + 8\,C_3E_3A_3 - 16\,A_3^{2}C_3  \\
	&& -4\,C_3B_3D_3+8\,C_3D_3A_3-4\,B_3E_3A_3, \\
	\Delta_4&=&
4\,B_3^{2}A_3+16\,C_3^{2}A_3-4\,C_3B_3^{2}-D_3^{2}C_3+E_3^{2}A_3+8\,C_3E_3A_3-16\,A_3^{2}C_3  \\
	&& -4\,C_3B_3D_3-8\,C_3D_3A_3+4\,B_3E_3A_3 .
\end{eqnarray*}
The expressions $g_2^{(2)}$ and $\Delta_5$ are  polynomials of degree
ten in the variables $A_3$, $B_3$, $C_3$, $D_3$ and $E_3$ (for instance,
$g_2^{(2)}$ contains 147 monomials).   $g_3^{(1)}$ is
a polynomial of degree 20 in the same variables.  Their explicit expressions
involve too many terms to be reproduced here.

Expressing  the coefficients $A_3, \cdots, E_3$ in terms of the entries
of the $R$-matrix, one  discovers further factorizations:
\begin{eqnarray*}
&&g_2^{(1)}=(c_1\,d_1-c_2\,d_2) \cdot  G_2^{(1)},  \qquad
\Delta_2=(c_1\,d_1-c_2\,d_2) \cdot  \delta_2,  \\
&&\Delta_3=(c_1\,d_1-c_2\,d_2)^2   \cdot \delta_3,  \qquad
\Delta_4=(c_1\,d_1-c_2\,d_2)^2  \cdot  \delta_4
\end{eqnarray*}
where $G_2^{(1)}$ is the sum of 1570  monomials of degree ten,
 $\delta_2$ is the sum of 104 monomials of degree six, $\delta_3$  and
$\delta_4$   are the sum of 780 monomials of degree eight, and
$\Delta_1$ is the sum of 256 monomials of degree eight in the entries
of the three-dimensional $R$-matrix, i.e. $a, \cdots, d_3$.

Let us note that $\Delta_2={\cal I}$ (see equation (\ref{I})), and is  thus
related to the projective invariant $\Ip$ of the conic (\ref{conic}).

\subsection{Subcases of \model}

The  three-dimensional model \model\
was built in such a way that it ``projects'' down to the two dimensional
Baxter model, as defined in section \ref{tr}.
It is natural  to consider the conditions on  \model\ obtained by writing that
the three projections
lie on the   critical or  disorder varieties of the Baxter model.

For example, writing the three disorder conditions
$a_B+d_B=b_B+c_B$~\cite{Ba81}
of the Baxter model  for the three projections ($i=1,2,3$),
yields  a codimension-three
subvariety of the three-dimensional model parametrized as follows:
\begin{eqnarray}
\label{deso3}
	&&a=b_1+b_2+b_3-2z, \qquad c_1=b_1+d_1-z \nonumber \\
	&&c_2=b_2+d_2-z, \qquad c_3=b_3+d_3-z
\end{eqnarray}
On the subvariety (\ref{deso3}), the discriminant of $P_4(s')$ vanishes,
$\Delta_2=\Delta_4=0$, the conic (\ref{conic}) degenerates since
$\Ip=\Delta_2=0$ and even more remarkably,
$P_4(s')$ gets proportional to $(s'-2)^4$.

Similarly, the three criticality conditions $a_B=b_B+c_B+d_B$~\cite{Ba81}
yield a codimension three subvariety:
\begin{eqnarray}
\label{crit3}
	2d_1 &=& a+b_1-b_2-b_3-2c_1, \qquad 2d_2=a+b_2-b_3-b_1-2c_2, \nonumber\\
	2d_3 &=& a+b_3-b_1-b_2-2c_3
\end{eqnarray}
On the subvariety (\ref{crit3}), the discriminant of $P_4(s')$ vanishes,
$\Delta_2=\Delta_3=0$, and now  $P_4(s')$ gets proportional to $(s'+2)^4$.
This last codimension-three subvariety is particularly interesting, since it
is $\Gamma_{3D}$ invariant.

\subsection{A solvable case.}
Another interesting model is obtained by setting  $d_1=d_2=d_3=0$.
The projections yield {\it six-vertex models}~\cite{LiWu72}.
The biquartic equation (\ref{can}) becomes a homogeneous equation of degree 4
and the solution is the union of four lines $r' = \lambda r$.
Remarkably, for this model the left-hand side of (\ref{can}) factorizes for
$r'=-r$:
\begin{eqnarray}
\label{factb}
	&& r^4 (ab_2+b_1 b_3-c_2^2+ab_1+b_2 b_3-c_1^2-2ac_3-2b_3c_3+2c_1c_2)
\nonumber \\
	&&\qquad (ab_2+b_1 b_3-c_2^2+ab_1+b_2 b_3-c_1^2+2ac_3+2b_3c_3-2c_1c_2) .
\end{eqnarray}

What is more interesting is that we can get the conditions for the existence
of solutions to (\ref{pt}) in this case.  In (\ref{pt}), there are two
equations which fix uniquely some scale factors:
\begin{equation}
	a = \mu, \qquad a\; pqr = \mu \;p'q'r'.
\end{equation}
Using these informations, the six others components of (\ref{pt}) fall into
three pairs of equations like:
\begin{eqnarray}
	a p' &=& b_1 p + c_3 q +c_2 r,
\nonumber\\
	a {1\over p'} &=& b_1 {1\over p} + c_3 {1\over q} +c_2 {1\over r}.
\end{eqnarray}
In writing these equations, we discarded the trivial solution $p=q=r=0$
which always exists in this case.
Multiplying pairwise these equations, we obtain three linear equations for
the variables $X_p=q/r+r/q$, $X_q=r/p+p/r$ and $X_r=p/q+q/p$.  This system of
equation can be easily solved.  These three variables are not independent
since they depend only on the ratios of $p$, $q$ and $r$.  They satisfy the
relation:
\begin{equation}
\label{xxx}
	X_p  X_q  X_r - (X_p^2 +X_q^2+X_r^2) +4 =0.
\end{equation}
Rewritten in the homogeneous variables $a$, $b_1$, $b_2$, $b_3$, $c_1$,
$c_2$, $c_3$, equation (\ref{xxx}) is a necessary condition for
equation (\ref{pt}) to have non-trivial solutions.  This is however not the
end of the story since the relation $pqr=p'q'r'$ yields another condition on
$R$ once we have solved for $r/p$ and $q/p$.  Note that the normalization of
the variables $p$, $q$, $r$ and $p'$, $q'$ and~$r'$ remains free.

A complete analysis therefore yields the existence of non trivial solutions
to (\ref{pt}) when $R$ is on some co\-dimension-two subvariety in the
parameter space.

\section{Conclusion.}

We have  shown how to  associate  algebraic curves with
three-dimensional vertex  models.  We have described a specific model
for which the analysis of these curves is handable.  We have introduced
a generalization of the quadratic Frobenius relations (associated to
elliptic functions).  It  corresponds to new intertwining relations of
products of more than two algebraic curves by $R$-matrices living on
algebraic varieties which are no longer curves. In the example detailed
in this paper, one has an intertwining of three curves by an $R$-matrix
living on a higher-dimensional algebraic variety.  We think that these
equations are a key ingredient for the construction of the
generalization of the Bethe Ansatz in higher dimensions, the quest of
solutions of the tetrahedron equations and more generally any exact
calculation (inversion trick~\cite{Ba80}, quest of critical
manifolds~\cite{WuWuBl89}) performed on higher dimensional models.

\vskip 1cm \noindent {\bf
Acknowledgement}: We thank J.~Avan and M.~Talon for many discussions and
encouragements.

\end{document}